\documentclass[twocolumn,showpacs,preprintnumbers,amsmath,amssymb,prl,superscriptaddress]{revtex4-1}
\usepackage[T1]{fontenc}
\usepackage[latin1]{inputenc}
\usepackage{graphicx}
\usepackage{amsfonts}
\usepackage{hyperref}

\newcommand{\buck}{{C$_{60}$}}

\newcommand{\siesta}{\textsc{Siesta}}

\newcommand{\beq}{\begin{equation}}
\newcommand{\eeq}{\end{equation}}

\newcommand{\beqa}{\begin{align}}
\newcommand{\eeqa}{\end{align}}

\newcommand{\etal}{\mbox{\textit{et al.}}}

\newcommand{\ev}{\mathrm{eV}}

\newcommand{\angstrom}{\mbox{\AA}}

\usepackage{xcolor}

\makeatother

\newif\ifchanged
\changedtrue
\changedfalse

\newcommand{\change}[1]{\ifchanged\textcolor{red}{#1}\else#1\fi}
\newcommand{\deleted}{\ifchanged\textcolor{red}{DELETE}\fi}

\begin{document}

\title{\change{Simple and efficient} LCAO basis sets for the diffuse states in carbon nanostructures}

\author{Nick R. Papior}
\author{Gaetano Calogero}
\author{Mads Brandbyge}

\affiliation{Dept. of Micro- and Nanotechnology, Technical University of Denmark, Center for Nanostructured Graphene (CNG), 
	{\O}rsteds Plads, Bldg.~345E, DK-2800 Kongens Lyngby, Denmark}

\email{nickpapior@gmail.com}

\date{\today}

\pacs{31.15.aq, 
    71.15.Mb, 
    31.15.E-, 
    31.50.Df, 
    71.15.Ap, 
    71.20.-b, 
}

\begin{abstract}
  We present a simple way to describe the lowest unoccupied diffuse states in carbon nanostructures
  in density functional theory (DFT) calculations using a minimal LCAO (linear combination
  of atomic orbitals) basis set. 
  By comparing plane wave basis calculations, we show how
  these states can be captured by adding long-range orbitals
  to the standard LCAO basis sets for the extreme 
  cases of planar \change{$sp^2$} (graphene) and \change{curved} carbon (\buck).  
  In particular, using Bessel functions with a long range
  as additional basis functions retain a minimal basis size.
  This provides a smaller and simpler atom-centered basis set compared to the standard
  pseudo-atomic orbitals (PAOs) with multiple polarization orbitals or by adding
  non-atom-centered states to the basis.
\end{abstract}

\maketitle

The bandstructure of graphene around the Fermi level is a textbook example of the
tight-binding model using just a single $p_z$-orbital ($z \perp$ graphene) per carbon atom
and nearest neighbour interaction\cite{CohenLouieBook}. Therefore it is not surprising
that it can be reproduced quite well by a simple linear combination of atomic orbitals
(LCAO) with a single orbital per valence state per carbon atom corresponding to one $s$
and three $p$ orbitals ($M=4$). It is generally of great interest to keep the basis-set
size ($M$) as small and simple as possible to keep the computational cost down, and to
enable calculations based on density functional theory (DFT) of larger systems, for
example electronic transport calculations of graphene-based devices using non-equilibrium
Greens functions\cite{Papior2015,Stradi2017}. An accurate description of all the occupied
bands, comparable to the result of calculations based on plane-waves (PW) basis sets,
calls for the use of a larger LCAO basis set f.ex. using $M=13$ \change{(two sets of $s$,
    $p$ and one set of $d$)} atom-centered basis
functions based on the atomic orbitals. This size is generally believed to be a good
compromise between accuracy and computational cost and is a standard choice in LCAO-DFT
codes such as \siesta\cite{Soler2002}, OpenMX\cite{open-mx}, or FHI-aims\cite{fhi-aims}.
However, as pointed out by Stewart\cite{Stewart2012}, this choice yields a wrong
description of the first unoccupied bands, which start about $3.25\,\mathrm{eV}$ above the
Fermi level and are parabolic around the Brillouin zone center, $\Gamma$. These bands
correspond to diffuse states with long tails into the vacuum, and are the first in a
quasi-continuum of free electron-like bands in a double Rydberg series of
image-potential-like states\cite{Silkin2009} with even and odd mirror symmetry in the
graphene plane. In particular, the first two unoccupied states ($1^\pm$) are important for
e.g. the description of interlayer states, reactivity,
intercalation\cite{Posternak1983,Agapito2016}, and tunneling into graphene, where the
inelastic phonon scattering plays a dominant role\cite{Zhang2008,Wehling2008a}.
States of similar origin has been found for the finite \deleted \buck-molecule, representing
another extreme compared to the flat, infinite graphene\cite{Feng2011,Hu2010}. The diffuse
molecular orbitals, dubbed Super Atom Molecular Orbitals (SAMOs) were observed in STM
experiments\cite{Feng2008}, and are located $\sim4\,\mathrm{eV}$ above the Fermi level.

Here we propose a simple, long-ranged, atomic-centered basis set, which can capture the
lowest unoccupied bands of graphene and the SAMO states of \buck\ in DFT-LCAO electronic
structure calculations. Its construction is based on a straightforward extension of
standard basis sets, and yield a level of accuracy comparable to PW calculations for the
first two unoccupied, diffuse bands (states) for graphene (\buck).

Quantum chemists traditionally use Gaussian-type orbitals(GTO) as bases. Another approach
is to use solutions to the free atoms, e.g. described by pseudo-potentials, and to confine
these within maximum range\cite{Sankey1989}. These pseudo-atomic orbitals (PAO) can be
used as a LCAO basis -- a so-called single-$\zeta$ (SZ) basis corresponding to $M=4$ for
carbon. The basis set can be improved by splitting each PAO into a part representing the
center and another part representing its tail, doubling the number of $\zeta$-functions
describing each valence orbital (double-$\zeta$ or DZ)\cite{Artacho1999}. To improve
further one can add orbitals with higher angular momentum ($l$) than present in the
valence shell, which for carbon amounts to the $d$-shell, $l=2$. These additional basis
functions are termed ``polarization'' and can be generated by applying a perturbing
polarizing electric field to the free atom\cite{Artacho1999}. The double-$\zeta$ plus
polarization (DZP) is thus amounting to $M=2\times 4 + 5=13$ basis functions for carbon
and comprise a standard LCAO basis set in \siesta\cite{Soler2002}, which is the DFT-LCAO
code we use in this study. However, one may use the splitting procedure\cite{Artacho1999}
to generate more refined bases such as triple-$\zeta$ or double-polarization. Importantly,
however these are all basis functions originating from an atomic problem and thus have a
decay away from the atomic core controlled by the atomic potential. We will return to this
point later, additionally our main discussion covers graphene while \buck\ SAMOs are
detailed in the end.

\begin{figure}
  \centering
  \includegraphics[width=\linewidth]{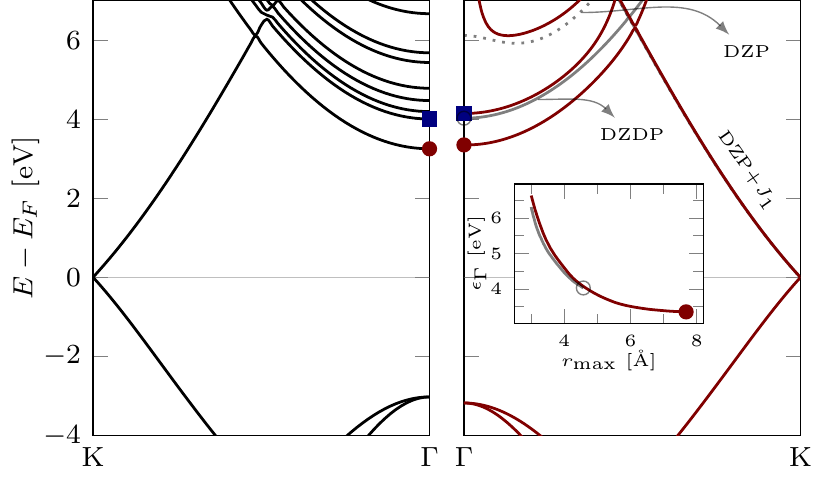}
  \caption{Bandstructure of graphene along $\Gamma$--K obtained from a PW basis set (left)
      in comparison with the equivalent bandstructure from three different LCAO basis sets
      (right): a standard DZP (grey dotted) with an orbital-confining radii cutoff
      $\delta E=0.0025\,\ev$, a DZDP (grey solid) with the same $\delta E$, and a DZP with
      $\delta E=0.1\,\ev$ combined with two Bessel functions $J_{l\in\{0,1\}}$ and hard-wall
      potential range $r_{\mathrm max}=7.5\,\angstrom$ (red solid). The eigenvalues at
      $\Gamma$ for the first ($1^+$) and second ($1^-$) lowest unoccupied bands are marked
      by a red circle and a blue square, respectively.  The insert shows the convergence
      of the $1^+$ eigenvalue at $\Gamma$ as a function of the basis ($J_l$) range for the
      DZDP (DZP+$J_1$) basis set. The maximum available DZDP range was constrained to
      $\sim 4.6\,\angstrom$ because of limitations in the choice of $\delta E$.
      \label{fig:bs}
  }
\end{figure}
Let us consider the electronic bandstructure of graphene in Fig.~\ref{fig:bs}, calculated
using a PW basis set from VASP\cite{Kresse1999} (left) and the DFT-LCAO code
\siesta\cite{Soler2002,zerothi_sisl} (right) for a selected choice of LCAO basis sets. We
have employed the PBE\cite{Perdew1996} functional for exchange-correlation, $k$-point
sampling of $42\times 42$ ($96\times 96$, LCAO), and a carbon-carbon distance of
$a=1.42\,\angstrom$. In the PW bandstructure the expected quasi-continuum of free
electron-like vacuum states appears at the $\Gamma$-point above $3.25\,\ev$. We focus on
the first ($1^+$) and second ($1^-$) lowest unoccupied eigenstates, marked by a circle and
square, respectively, and compare them to the LCAO bandstructures. We consider three
different atom-centered bases, namely the standard DZP ($M=13$), a double-polarization
DZDP ($M=18$) where the polarization $d$-orbitals are doubled, and a DZP basis extended by
two Bessel functions ($J_l$) with angular momentum $l\in\{0,1\}$ ($M=13+4=17$), in the
following $J_1$ implicitly includes $J_0$ orbitals.

First we note that while all LCAO bases yield a good description of the occupied bands,
the standard DZP basis set fails completely in reproducing the lowest unoccupied states,
showing a non-parabolic $1^+$ band around $\sim 5.8\,\mathrm{eV}$ at $\Gamma$. The results
suggests that the discrepancy is due to the limited DZP basis size, which cannot supply
linear combinations to account for the free electron-like bands. As shown by Silkin
\etal\cite{Silkin2009} the $1^\pm$ bands have $s$ and $p_z$ characters which are already
the predominant part of the valence bands. \deleted Therefore the DZP basis can
not account for \emph{both} the free electron-like bands \emph{and} the valence bands.
%

The easiest procedure towards correcting the shape and position of the lowest unoccupied
band is to double the polarization orbitals, DZDP (or TZDP \cite{Agapito2016}). In this
case it is the tail polarization $d$-orbitals that accounts for the missing linear
combinations. Subsequent tuning of the range of the basis is necessary in order to obtain
a better agreement with the PW results. This is done in the inset of Fig.~\ref{fig:bs}
which shows the convergence of the DZDP (gray) lowest band at $\Gamma$ ($\epsilon_\Gamma$)
with respect to the basis orbital cutoff radius.


We can obtain better and more economical results by using custom basis orbitals based on
spherical Bessel functions\cite{Haynes1997}. The Bessel functions are solutions to the
spherical ``particle-in-a-box'' problem with hard-wall
cutoff\cite{Griffiths}. Importantly, these orbitals are not constrained by a core
potential, and thus have a well defined shape depending \emph{only} on the chosen radial
cut-off
and angular momentum $l$.
An atomic orbital does not necessarily increase weight for large $r$ due to confinement
potentials. Effectively this means that basis orbitals originating from atomic
pseudopotentials tend to have a small cutoff radius regardless of user defined ranges.
The first band can be described by a single long range $J_0$ Bessel function ($s$), while
the second band also requires $J_1$ ($s+p$). The \mbox{$\Gamma$--K} bandstructure in
Fig.~\ref{fig:bs}(right) shows the DZP$+J_1$ which is in good agreement with the PW
calculation for the first two bands. An improved energy alignment with respect to the PW
calculation can be achieved by extending the basis orbitals to as much as
$r_{\mathrm{max}}\sim7.5\,\angstrom$ which was used above. The inset in Fig.~\ref{fig:bs}
shows the convergence of the first band energy at the $\Gamma$-point for increasing
$r_{\mathrm{max}}$ for the DZDP and DZP$+J_1$ basis sets.

Remark that DZ$+J_0$/DZ$+J_1$/DZP$+J_0$/DZP$+J_1$ all reproduce the first band with a band
onset between $3.41\,\ev$ and $3.35\,\ev$, respectively. In Table~\ref{tab} we list the
two first unoccupied band-onsets at $\Gamma$ for the prominent DZ variants tested. All SZ
variants yield $\epsilon_{1^+}>5\,\ev$, while the TZ variants are comparable to DZ.
\begin{table}[!h]
  \footnotesize
  \caption{Positions of band energies at the $\Gamma$-point for the first ($1^+$) and
      second ($1^-$) unoccupied bands for different LCAO basis sets with size $M$ and
      maximal cut-off. PW
      shows the planewave benchmark calculation.}
  \label{tab}
  \let\mysize\scriptsize
  \setlength{\tabcolsep}{1.6pt}
  \begin{tabular}{c|ccccccc|c}
  $[\ev]$  & \mysize DZ &\mysize DZ$+J_0$ &\mysize DZ$+J_1$ &\mysize DZP
    &\mysize DZP$+J_0$
    &\mysize DZP$+J_1$ &\mysize DZDP & \mysize PW
    \\
    \hline
    $\epsilon_{1^+}\includegraphics{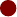}$ &
    $9.03$ & $3.42$ & $3.35$ & $5.81$ & $3.41$ & $3.35$ & $3.64$ & $3.25$
    \\
    $\epsilon_{1^-}\includegraphics{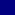}$ &
    $9.33$ & $9.34$ & $4.07$ & $8.24$ & $8.30$ & $4.15$ & $8.27$ & $4.01$
    \\
    $M$ & $8$ & $9$ & $12$ & $13$ & $14$ & $17$ & $18$ & ---
  \end{tabular}
\end{table}


We note in passing that one could include the unoccupied band (only close to $\Gamma$) in
the nearest neighbour tight-binding $p_z$-model\cite{CohenLouieBook} by adding an orbital
with $s$-symmetry to each atom, thus yielding orbitals orthogonal to the $\pi$-system. The
hopping parameter $\gamma^+$ can be approximated by the regular $p_z$ hopping parameter
since the two bands have nearly identical parabolic curvature close to $\Gamma$,
effectively setting $\gamma^+ \approx 2.7\,\ev$ and on-site $\epsilon_{1^+} +
3\gamma^+$. Further discussion of tight-binding models of the bands may be found
elsewhere\cite{PhysRevB.89.165430}.

\begin{figure}
  \centering
  \includegraphics[width=\linewidth]{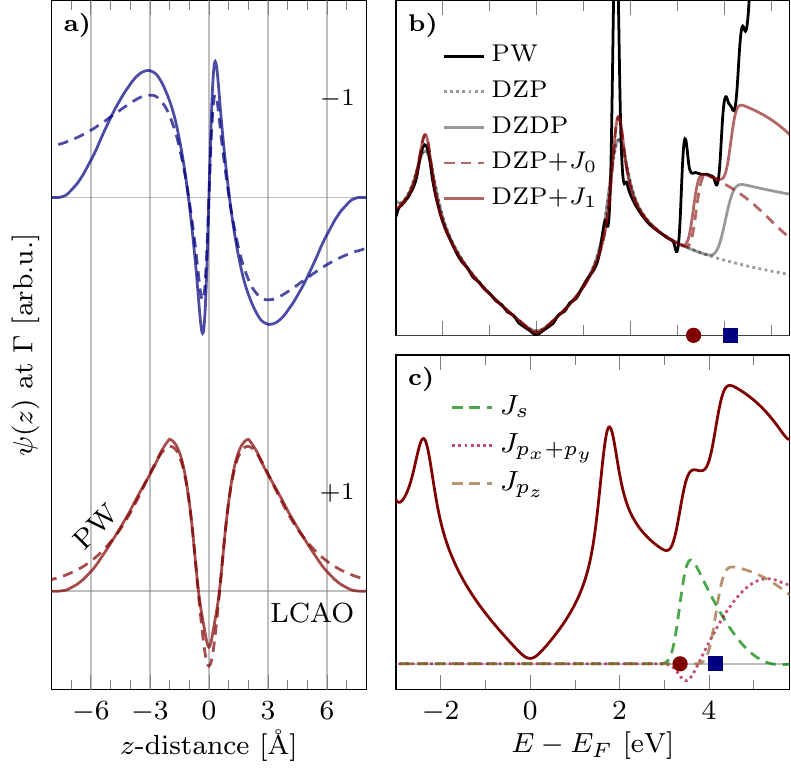}
  \caption{ (a) Comparison of normalized  wavefunctions at $\Gamma$ obtained with PW basis
      (dashed), and LCAO with the DZP+$J_1$ basis (full). The wavefunctions at $\Gamma$
      are projected on a line through a carbon atom. \change{The $+1/-1$ states are plotted in
      bottom/top parts, respectively.} (b) Total DOS from PW in comparison
      with DZP, DZDP and DZP$+J_0$/$J_1$ basis sets. The onsets of $1^\pm$ at $\Gamma$ are
      highlighted on the energy axis.  (c) Orbital resolved highest contributions to the
      DOS from DZP+$J_1$ in correspondence of the two lowest unoccupied bands. \deleted
      \label{fig:wf}
  }
\end{figure}

In Fig.~\ref{fig:wf}a we compare the wavefunctions through a carbon atom along $z$
obtained by PW and LCAO, respectively. These also show a reasonable agreement with the PW
results. Note how the LCAO tails are forced zero for $r>7.5\,\angstrom$. The symmetric
lowest state $1^+$ (bottom) is accurately described by LCAO although the tail for PW
extends farther into vacuum. The anti-symmetric second lowest state $1^-$ (top) is more
extended in PW compared to LCAO, as expected.

The density of states (DOS) is shown in Fig.~\ref{fig:wf}b comparing the PW calculation
with the four selected basis sizes. $k$-point sampling was converged. A large improvement
in the description of the unoccupied bands accompanies the appropriate choice of basis
size. Clearly, DZP$+J_0$/$J_1$ reproduce the DOS to a satisfactory level.  The difference
between PW and LCAO DOS shapes are mainly due to different smearing methods.  In
Fig.~\ref{fig:wf}c the projected DOS onto the basis functions for DZP$+J_1$ highlights how
the unoccupied bands indeed are of $s$- ($1^+$) and $p$-character ($1^-$).  Thus the
$1^+$-state consists of $s$ with a negative $p_{x(y)}$ where the band starts (at
$\Gamma$), while the $1^-$-state has $p_z$ odd character symmetry.


Along similar lines Agapito and co-workers\cite{Agapito2016} considered projections of
different LCAO basis-sets onto plane-wave Bloch states as well as DFT-LCAO
calculations. They found that a TZDP ($M=22$) basis set with a cutoff range of
$4\,\angstrom$ did not reproduce the $1^\pm$-bands and had to use a DZP supplemented with
long-ranged (cutoff $6.9\,\angstrom$) empty-atom (EA) basis-functions located
$2.8\,\angstrom$ outside the graphene plane to get a reasonable description of
these. Besides being costly to use a DZP+EA basis, it also makes calculations conceptually
and practically more difficult for systems where one e.g. adsorb or bind molecules to
graphene. A large overlap between the EA-basis and the adsorbates may lead to spurious
effects.

\begin{figure}
  \centering
  \includegraphics{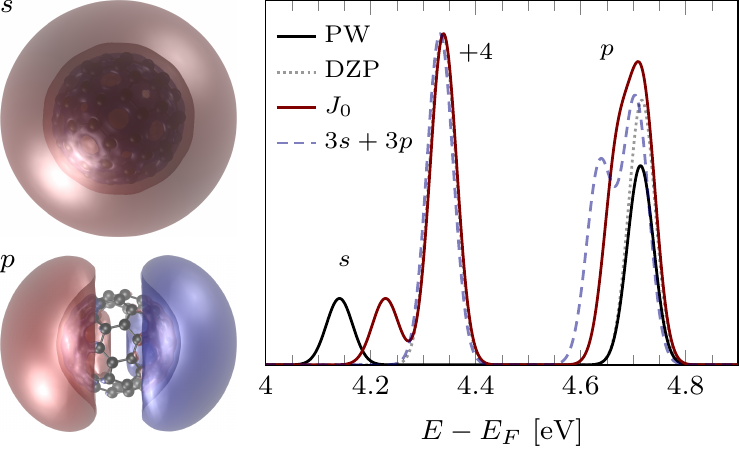}
  \caption{Left: Iso-surface plots of the $s$/$p$ SAMO from LCAO $J_0$ calculation.  %
      Right: DOS comparison with PW, DZP, $J_0$ and $3s+3p$ basis\change{, aligned at
          $E_F$}. \change{The LUMO$+4$ level is indicated as a reference to the rest of
          the \buck-states. The $J_0$ basis reproduces the $s$ and $p$ SAMO while the much
          larger $3s+3p$ basis only reproduces the $p$ SAMO.}
      \label{fig:samo}
  }
\end{figure}

As outlined above Bessel functions are advantageous in the graphene case. Another approach
uses long range $3s$ and $3p$ carbon atomic orbitals, which also correctly describes the
graphene unoccupied states and with equivalent precision and basis size $M$ as
$J_1$. However, for SAMO states of the \buck-molecule we could only reproduce the
$s$-character SAMO using the Bessel basis (further fine tuning of $3s/3p$ orbitals may be
able \change{to} capture the SAMOs). In Figure~\ref{fig:samo} we show the wavefunctions of
the $s$ and $p$ SAMOs (produced by DZP$+J_0$) along with the DOS in the respective energy
range, PW calculations using $(40\,\angstrom)^3$ cell. The shape of the wavefunctions
compare well with those obtained with PW calculations\cite{Feng2008}. Comparing DOS shows
that the $s$ SAMO for PW and $J_0$ are separated by $\sim 0.1\,\mathrm{eV}$, while the
$3s+3p$ could not reproduce the $s$ state. \change{Note that the PW LUMO$+4$ position is
    not in this energy range. These states are highly dependent on the cell (vacuum)
    size. The PW $s$ SAMO state is fixed in energy for $30\,\angstrom$ and
    $40\,\angstrom$ cell sizes.}

In conclusion we have shown that the two lowest unoccupied diffuse states for graphene and
\buck\ can adequately be described within the DFT-LCAO framework by adopting a
conceptually and computationally simple atomic-centered basis set where Bessel functions
with a long extension are supplementing the standard DZP-basis. The presented basis set
provides a good compromise with respect to efficiency, due to the relatively small number
of orbitals required, while ensuring a level of accuracy which is comparable to DFT
calculations based on the planewave basis. \change{The Bessel basis sets may be relevant in
    other 2D materials and/or surface calculations\cite{Moreno2018,Garcia-Gil2009}.} The
first two bands of graphene may be selected by choosing the symmetry of the basis function
($J_0$ or $J_1$), while for \buck\ $J_0$ is enough. Consequently only adding $1$ basis
orbital per atom to the DZ/DZP basis set will correctly describe the first unoccupied band
of graphene.

We are grateful to Dr. Aran Garcia-Lekue and Mr. Bernhard Kretz for useful discussions.
We acknowledge funding from Villum Fonden (grant no. 00013340), Danish research council
(grant no. 4184-00030). The Center for Nanostructured Graphene (CNG) is sponsored by the
Danish Research Foundation, Project DNRF103.

\bibliography{article}

\end{document}